\documentclass[letterpaper]{sig-alternate}
\usepackage{amssymb,booktabs,multirow,algorithm,algorithmic,xcolor ,amsmath,url}
\usepackage{subfigure}
\usepackage{graphicx}
\usepackage{balance}
\usepackage{wrapfig}
\usepackage{graphicx}
\usepackage{picins}
\usepackage{color}
\permission{\copyright 2017 International World Wide Web Conference Committee (IW3C2), 
published under Creative Commons CC BY 4.0 License.}
\conferenceinfo{WWW'17 Companion,}{April 3--7, 2017, Perth, Australia.}
\copyrightetc{ACM \the\acmcopyr}
\crdata{978-1-4503-4914-7/17/04. \\
http://dx.doi.org/10.1145/3041021.3054722 \\
\includegraphics{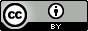}
}
 
\begin{document}
\title{TruthDiscover: Resolving Object Conflicts on Massive Linked Data}
%
%
%
%
%

\numberofauthors{1} 
%
\author{
%
%
\alignauthor
Wenqiang Liu$^{1}$, Jun Liu$^{1}$, Haimeng Duan$^{1}$, Jian Zhang$^{1}$, Wei Hu$^{2}$, and Bifan Wei$^{1}$
\newline\newline
\affaddr{$^{1}$MOEKLINNS Lab, Department of Computer Science, Xi'an Jiaotong University, China}\\
\affaddr{$^{2}$State Key Laboratory for Novel Software Technology, Nanjing University, China}\\
\email{liuwenqiangcs@gmail.com, liukeen@mail.xjtu.edu.cn, duanhaimeng@gmail.com zjxzj520@gmail.com, whu@nju.edu.cn, weibifan@mail.xjtu.edu.cn}
}

\maketitle
\begin{abstract}
Considerable effort has been made to increase the scale of Linked Data. However, because of the openness of the Semantic Web and the ease of extracting Linked Data from semi-structured sources (e.g., Wikipedia) and unstructured sources, many Linked Data sources often provide conflicting objects for a certain predicate of a real-world entity. Existing methods cannot be trivially extended to resolve conflicts in Linked Data because Linked Data has a scale-free property. In this demonstration, we present a novel system called TruthDiscover, to identify the truth in Linked Data with a scale-free property. First, TruthDiscover leverages the topological properties of the Source Belief Graph to estimate the priori beliefs of sources, which are utilized to smooth the trustworthiness of sources. Second, the Hidden Markov Random Field is utilized to model interdependencies among objects for estimating the trust values of objects accurately. TruthDiscover can visualize the process of resolving conflicts in Linked Data.
\end{abstract}



\section{Introduction}
As of August 2014, the number of available Linked Data sources has increased from 12 in 2007 to 1,014 with more than 4 billion RDF triples from a variety of domains \cite{schmachtenberg2014state}. Given that most Linked Data sources have been created from semi-structured sources (e.g., Wikipedia) and unstructured sources (e.g., text)\cite{dutta2014probabilistic}, conflicting objects from multiple Linked Data sources for a certain predicate of a real-world entity become inevitable. For example, Freebase\footnote{https://www.freebase.com/} and Yago \cite{fabian2007yago} provide different values for the predicate \textit{dbp:height} of \textit{Statue of Liberty}. The problem of object conflicts has negative impact to developers aiming to seamlessly consume and integrate Linked Data in their applications. Hence, this problem must be addressed.

\begin{table}[htp]
 \centering
\caption{Partly predicates of \textit{Statue of Liberty}.}
	\begin{tabular}{*{3}{c}}
	\hline
	\multicolumn{1}{c}{\multirow{2}{*}{Sources}}
  &\multicolumn{2}{c}{Predicates}
  \\
   \cline{2-3}
   \multicolumn{1}{c}{}
   &\multicolumn{1}{c}{dbp:height} &{dbp:beginningDate}
   \\
      \hline\hline
	DBpedia & NULL & 1886-10-28 \\
	Freebase & 93 & 10/28/1886 \\
	Yago & 46.0248 & 1886-\#-\# \\
	Wikidata & NULL & 28 October 1886  \\
	\hline
	\end{tabular}
\end{table}
  
A straightforward method to resolve object conflicts is majority voting, where the object with the maximum number of occurrences is regarded as truth. However, we find that this method achieves relatively low accuracy (ranging from 0.3 to 0.45) in Linked Data because many predicates have no dominant object. To address the limitation of the majority voting, many methods based on truth discovery have been proposed \cite{li2014resolving,yin2008truth}. A common principle of these methods is that a source which provides trustworthy objects more often is more reliable, and an object from a reliable source is more trustworthy. However, the effectiveness of existing truth discovery methods is significantly affected by the number of objects provided by each source. In our previous work \cite{liu2015truth}, we found that the number of conflicting objects provided by most of the sources ranges from 1 to 10, and only a few sources have many conflicting objects. This finding indicates that Linked Data has a scale-free property. Therefore, these methods cannot be trivially extended to resolve conflicts in Linked Data.

In this study, we developed a novel system called TruthDiscover\footnote{ A  introduction at https://youtu.be/TtnUNl87FVU.} that can reduce the effect of the scale-free property on truth discovery. The following are the key features of TruthDiscover.

1) TruthDiscover leverages the topological properties of the Source Belief Graph (see Section 2 for the definition of \textit{Source Belief Graph}) to estimate the priori beliefs of sources for smoothing the trustworthiness of sources.

2) The Hidden Markov Random Field is utilized in this system to model interdependencies between objects for estimating the trust values of objects accurately.

3) TruthDiscover provides a graphical interface to visualize the process of resolving objects conflicts for a certain real-world entity.

The remainder of this demonstration is organized as follows. Several concepts about TruthDiscover are discussed in Section 2. Section 3 shows the architecture of TruthDiscover and explains important technical issues. Section 4 demonstrates how the TruthDiscover offers a graphical interface and reports the preliminary results on four domains. Section 5 presents the conclusions of this demonstration.
\section{PRELIMINARIES}
Several important notations utilized in TruthDiscover are introduced in this section.

\textbf{SameAs Graph:} Given a set of RDF triples $T$, each RDF triple can be represented by $\langle  s,p,o \rangle$, where $s$ is a subject, $p$ is a predicate, and $o$ is an object. A SameAs Graph can be represented by $SG=(V,E)$, where $V=\{s|\langle s,\text{owl:sameAs},o\rangle \in T\}\cup\{o|\langle s,\text{owl:sameAs},o \rangle \in T \} $  is a set of vertices, $E\subseteq V \times V$ is a set of directed edges with each edge corresponding to an triple in $T$.

\textbf{Source Belief Graph} \cite{liu2015truth}\textbf{:} Given a SameAs Graph $SG$, the Source Belief Graph can be denoted by $SBG=(\mathcal{W},R)$, where   $\mathcal{W}$ is a set of vertices with each vertex corresponding to the source name of the vertex in SameAs Graph $SG$; $R$ is a multiset of $\mathcal{W}  \times \mathcal{W}$ formed by pairs of vertices $(\mu,\nu)$, $\mu,\nu \in \mathcal{W}$ and each pair $(\mu,\nu)$ corresponds to an edge in SameAs Graph $SG$.

\textbf{Trustworthiness of Sources} \cite{li2014resolving}\textbf{:} The trustworthiness of a source $\omega_j $ is the expected confidence of the objects provided by $\omega_j $, denoted by $t(\omega_j) $.

\textbf{Trust Values of Objects} \cite{li2014resolving}\textbf{:} The trust value of an object $o_i $ is the probability of being correct, denoted by $\tau(o_i) $.

We let $O\text{=}\{o_{i}\}_{m}$ denote a set of conflicting objects for a certain predicate of a real-world entity. The process of resolving object conflicts in Linked Data is formally defined as follows: given a set of conflicting objects $O$, TruthDiscover will produce one truth for a certain predicate of a real-world entity. The truth is represented by $o^*=\arg\underset{o_{i}^{}\in{O}}{\max}  \; \tau(o_{i})$.
\section{TruthDiscover ARCHITECTURE}
TruthDiscover will produce one truth for each predicate that have conflicting objects by employing a three-modules framework, as shown in Figure 1.

\begin{figure}[!htb]
\flushleft
\includegraphics[width=0.48\textwidth]{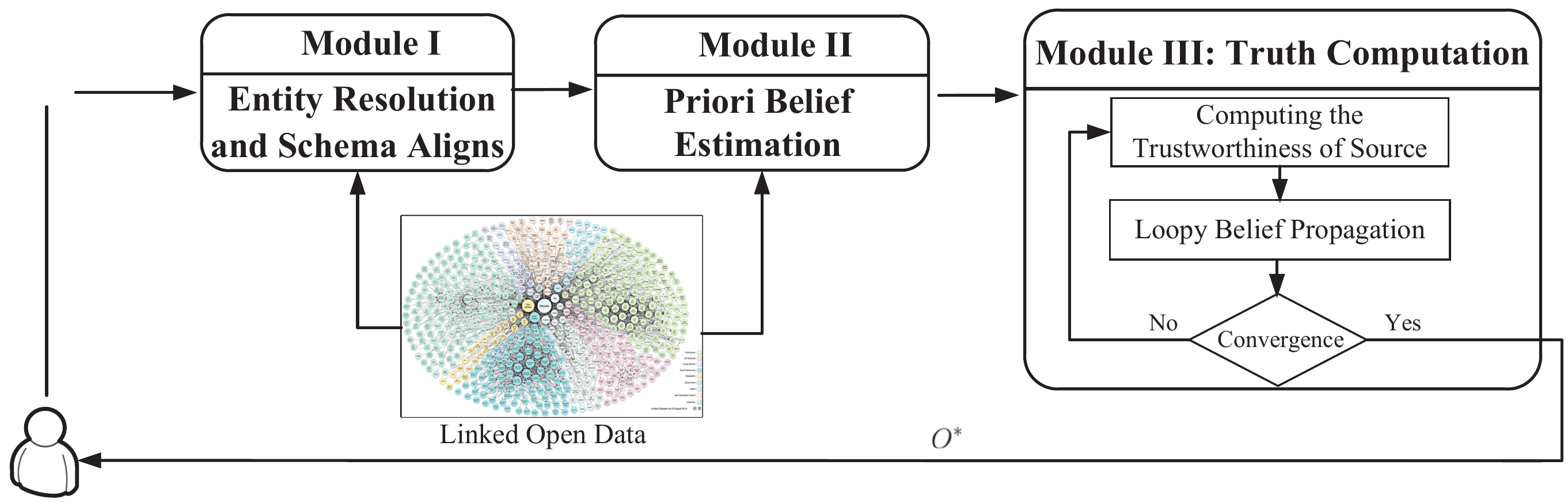}
\caption{Framework of TruthDiscover}
\end{figure}

 \textbf{Module I. Entity Resolution and Schema Alignment:} Firstly, we perform entity co-reference resolution through the API of sameas.org\footnote{http://sameas.org/}, which is a well-known tool, to identify subjects for the same real-world entities. Then, a method based on supervised learning is used to schema mapping. 

 \textbf{Module II. Priori Belief Estimation:} In this module, the priori belief of each source  are produced by leveraging the topological properties of the Source Belief Graph in order to reduce the effect of the scale-free property on truth discovery.

 \textbf{Module III. Truth computation}: Firstly, the trustworthiness of each source is automatically computed based on the trust scores of objects and the priori beliefs of sources. Therefore, the loopy belief propagation algorithm is applied to estimate the marginal probabilities of each hidden variable in HMRF. If the changes in all objects after each iteration are less than the threshold, then the object with the highest trust score is regarded as the truth.
 
Module I crawls Linked Data and completes the resolution of schema for a real-world entity, which are not focus of this paper. so the detailed implementation of the crawling algorithm will not be described further. In the following section, we will only focus on the description of modules II and III.
\subsection{Priori Belief Estimation}
Module II automatically estimates the priori belief $BR(\omega_j)$ of source $\omega_j $ by leveraging the topological properties of the Source Belief Graph $SBG$.

As we all know, the \textit{owl:sameAs} property in Linked Data indicates that two subjects actually refer to the same thing. When data publishers publish their data as Linked Data on the web, they add new owl:sameAs triples pointing to the external equivalent subject~\cite{bizer2009linked}. As dictated by logic, the owl:sameAs property indicates that the data publishers place their focus and trust to the subject provided by a source they trust.

When many of \textit{owl:sameAs} triples are taken together, they form a directed graph called SameAs Graph. The SameAs Graph can be converted to a directed multigraph called the Source Belief Graph, which represents the relationship between sources. The Source Belief Graph indicates that the trustworthiness of different sources can be propagated through the edges. The edge structure of the Source Belief Graph is utilized to produce a global reliability ranking of each source. The priori belief $BR(\omega_j)$ of source $\omega_j $ can be defined as follows: 
\begin{equation}
BR(\omega_j)=(1-d)+d*\sum_{\omega_l \in B_{\omega_j}} \frac{BR(\omega_l) L(\omega_l,\omega_j)}{C(\omega_l)}
 \end{equation}
 where parameter $B_{\omega_j} $ denotes the set of sources that point to $\omega_j $; $C(\omega_j) $ denotes the number of edges going out of source $\omega_j $; $L(\omega_l,\omega_j) $ presents the number of edges that $\omega_l $ point to $\omega_j $ and $d $ is a damping factor.
 
 Figure 2 shows a screenshot of the module II that illustrates the procedure of priori belief estimation.
 
 \begin{figure*}[!htb]
\begin{minipage}[t]{0.49\textwidth}
\centering
\includegraphics[width=0.95\textwidth]{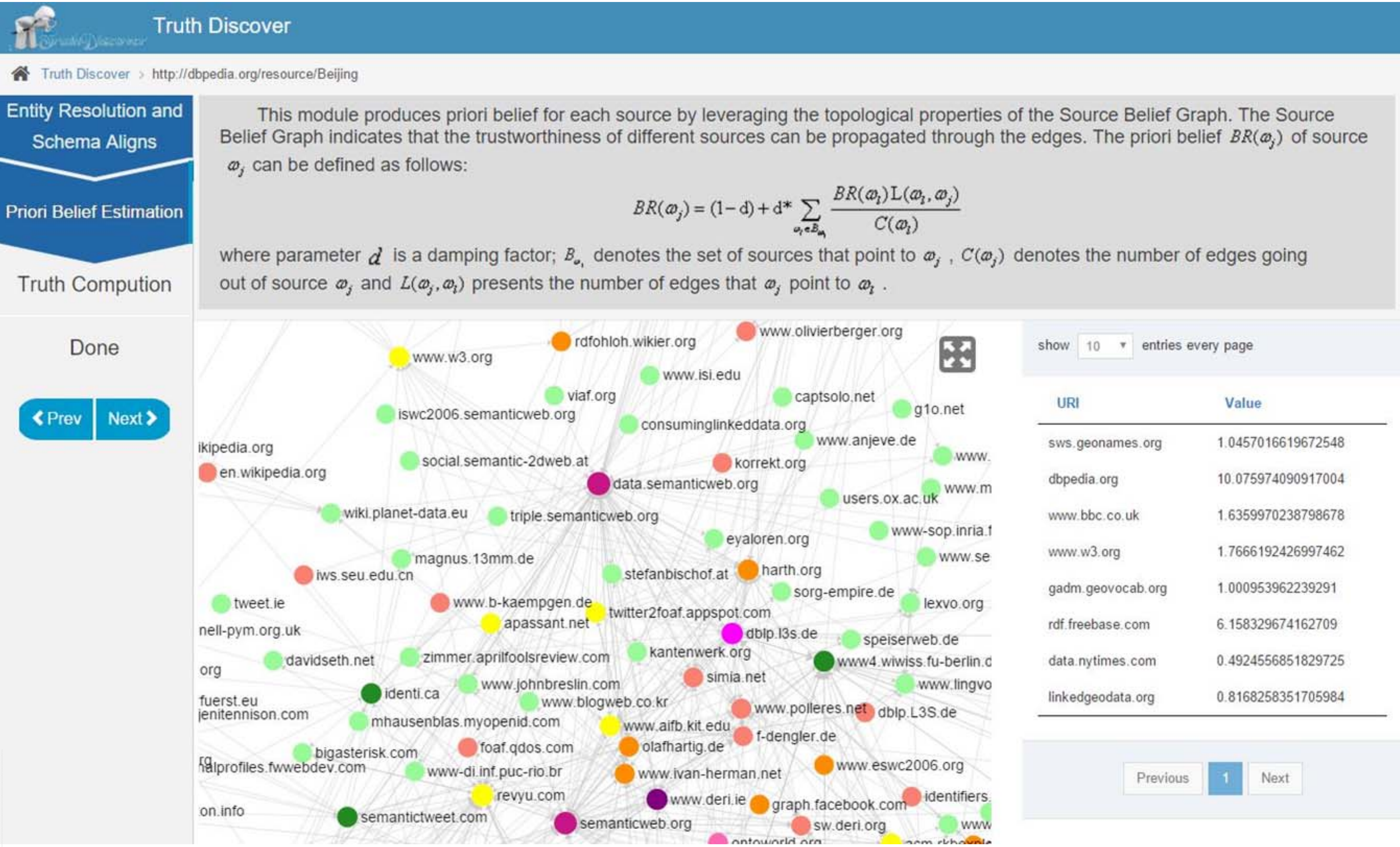}
\caption{Screenshot of module II.}
\label{object}
\end{minipage}%
\begin{minipage}[t]{0.49\textwidth}
\centering
\includegraphics[width=0.95\textwidth]{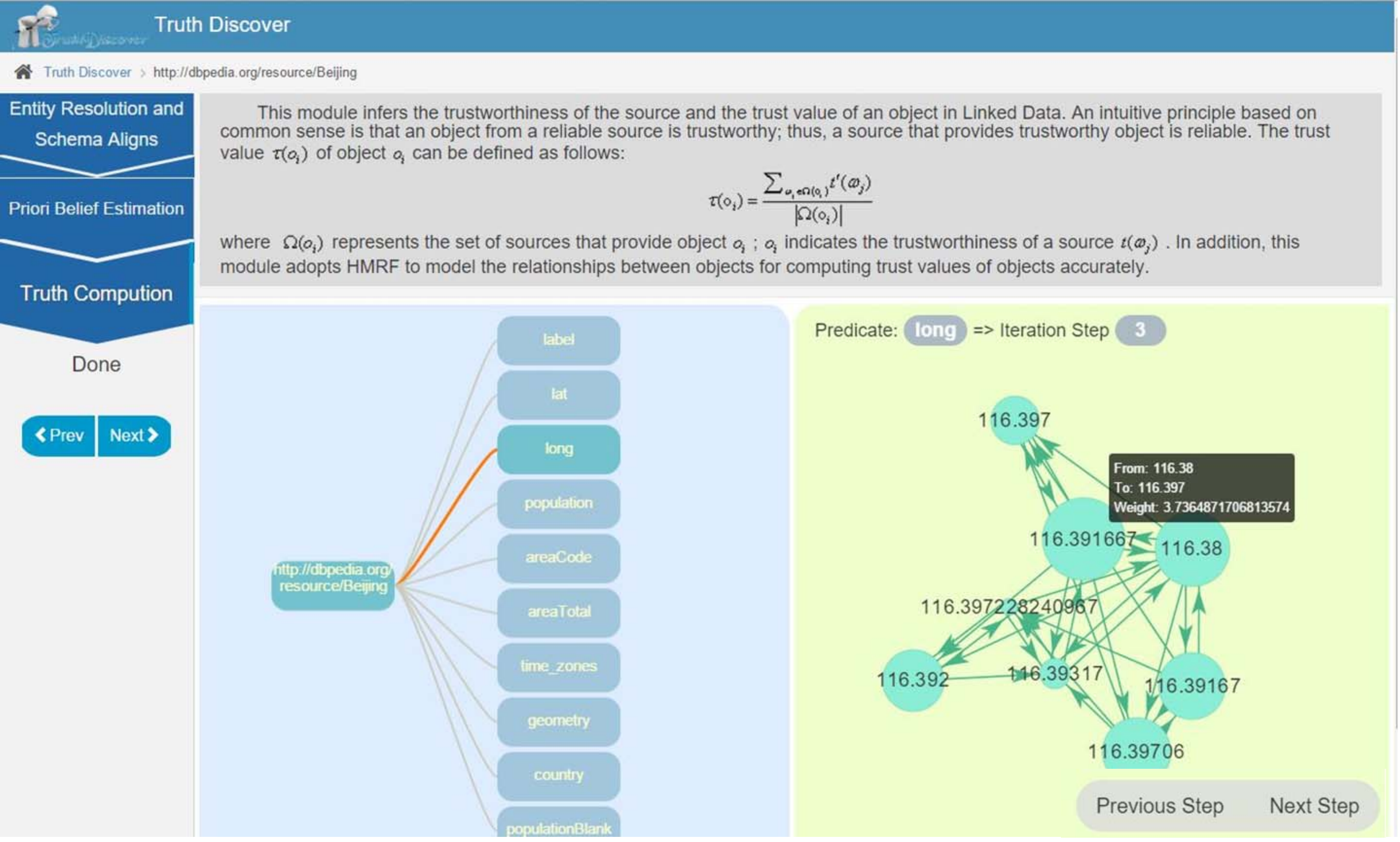}
\caption{Screenshot of module III.}
\label{object}
\end{minipage}
\end{figure*}

\subsection{Truth computation}
Module III infers the trustworthiness of the source and the trust value of an object in Linked Data with a scale-free property. The computation of trustworthiness of the source and the trust value of an object can be further divided into three steps.

\textbf{Step 1. Computing the Trustworthiness of Sources:} In this paper, the trustworthiness $t(\omega_j) $ of a source $\omega_j $ is regarded as the average probability of the object provided by $\omega_j $ being true as defined as follows:
\begin{equation}
t(\omega_j)=\frac{\sum_{o_i \in F(\omega_j)}\tau(o_i)}{|F(\omega_j)|}
 \end{equation}
 where $F(\omega_j)$ is the set of objects provided by source $\omega_j $.
 
Considering the scale-free property of Linked Data, it's difficult for Equation 2 to estimate the real reliability degree of source $\omega_j $ accurately when$|F(\omega_j)|$ is ``small." In this study, the trustworthiness  $t(\omega_j)$ of source $\omega_j $ is smoothed by priori belief $BR(\omega_j)$ based on the averaging strategy as defined as follows: 
\begin{equation}
t'(\omega_j)=\frac{NBR(\omega_j)+t(\omega_j)}{2}
 \end{equation}
 \begin{equation}
NBR(\omega_j)=\frac{BR(\omega_j)-min}{max-min}
 \end{equation}
 where $NBR(\omega_j)$ represents the normalized priori belief of $\omega_j $; $max$ and $min$ indicate the maximum and minimum values of all priori beliefs respectively.
 
\textbf{Step 2. Computing the Trust Values of Objects:} First, the trust value $\tau(o_i) $ of object $o_i $ can be defined as follows:
 \begin{equation}
\tau(o_i)=\frac{\sum_{\omega_j \in \Omega (o_i)}t'(\omega_j)}{|\Omega (o_i)|}
 \end{equation}
where $\Omega(o_i)$ represents the set of sources that provide object $o_{i}$.

Second, TruthDiscover exploits following two findings in order to estimate the trust values of objects more accurately.\\
\textbf{Finding 1:} The true objects appear to be similar in different sources.\\
\textbf{Finding 2:} The false objects are less likely to be similar. \\
These two findings indicate that the trust value of an object can propagate to other objects through the similarity relation. TruthDiscover models the relationship between objects by adopting a method based on HMRF. We let the observation variables $O\text{=}\{o_i\}_{m}$ are a set of conflicting objects for a certain predicate of a real-world entity. The hidden variables $
Y=\{y_i\}_m $ are the labels of $o_i$. Each hidden variable $y_i\in \{0,1\}$ indicates whether corresponding object is a truth. Therefore, the joint distribution of variables in HMRF is factorized as follows:
 \begin{equation}
P(y_1,y_2,...,y_m)\text{=}\frac{1}{Z}\prod_{c \in \rm C} \psi_c(X_c)
 \end{equation}
 \begin{equation}
Z\text{=}\sum_{X_c \in X} \prod_{c \in \rm C} \psi_c(X_c)
 \end{equation}
 where $Z$ is a constant selected to ensure that the distribution is normalized, $\rm C$ denotes the set of all maximal cliques, $X_c$ indicates the the set of variables of a maximal clique and $\psi_c(X_c)$ is a potential function in HMRF.
 
 In order to estimate the marginal probabilities of hidden variable, the belief propagation algorithm is adopted in TruthDiscover. $\tau(o_i) $ will converge after a sufficient number of iterations.
 
 \textbf{Step 3. Iterative computation:} Because of the interdependencies between the trustworthiness of sources and the trust value of objects, TruthDiscover adopt iterative strategy to infer the trustworthiness of sources and trust values of objects. In each step of the iterative procedure, TruthDiscover first uses the trustworthiness of sources to compute trust values of objects and then the recomputes the trustworthiness of sources. If the changes in all objects after each iteration are less than the preset threshold, then object with the maximum trust score is regarded as the truth.
 
\section{DEMONSTRATION DETAILS}
We provide an interactive UI to demonstrate the effectiveness of TruthDiscover on large-scale real RDF datasets.
\subsection{Demonstration Setup}
TruthDiscover is implemented as a Java Web application. It allows users to search their interested subject(e.g., http://dbpedia.org/resource/Beijing) via a Web-based interface. To visualize the Source Belief Graph and process of iterative computation, Sigmajs\footnote{http://sigmajs.org/} (an open-source tool for integrating network exploration in Web applications) and JavaScript InfoVis Toolkit\footnote{http://philogb.github.io/jit/} (an open-source tool for creating interactive data visualizations on the Web applications) are incorporated into TruthDiscover for graph/network analysis and visualization. 
\subsection{Experimental Evaluation}
Four datasets that belong to four domains: \textit{people}, \textit{locations}, \textit{organizations} and \textit{descriptors} were constructed based on the OAEI2011 New York Times dataset\footnote{http://data.nytimes.com/} and BTC2012\footnote{https://km.aifb.kit.edu/projects/btc-2012/}, which are two well-known and carefully created datasets of Linked Data. The statistics of the four datasets are shown in Table 2.
\begin{table}[htp]
	\caption{Statistics of the Four datasets.}
	\centering
	\begin{tabular}{cccc}
		\hline
Domains &\#Subjects &\# Predicates &\# Conflicting  \\
&&&Predicates\\
	\hline\hline		
 Person & 130174& 16245& 7506\\
 Locations& 74015& 14162& 6870\\
 Organizations& 25051& 13956& 6360\\
 Descriptors& 10362& 6980& 3250\\ 
 \hline
	\end{tabular}
\end{table}

We select three well-known state-of-the-art truth discovery methods as baseline including Vote, TruthFinder~\cite{yin2008truth} and F-Quality Assessment \cite{michelfeit2014linked}. The parameters of the baseline methods are set according to the authors's suggestions. The experiments are performed on a desktop computer with Intel Core i5-3470 CPU 3.2 GHz with 4 GB main memory, and Microsoft Windows 7 professional operating system. All baseline methods were executed in the Eclipse (Java) platform  by a single thread. Figure 4 shows the experimental results of all the methods in terms of accuracy in the four datasets.

 \begin{figure}[!htb]
\centering
\includegraphics[width=0.5\textwidth]{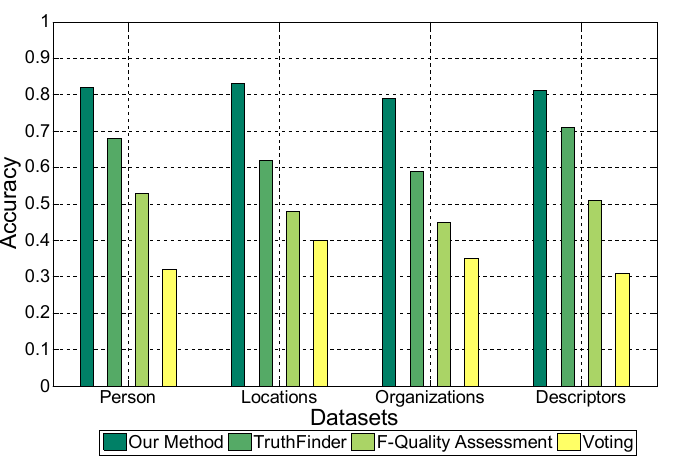}
\caption{Performance comparison in the Four datasets}
\end{figure}

It can be concluded from Figure 4 that TruthDiscover outperforms the three baseline methods in terms of accuracy. The main reason for this superiority is that it's difficult for three baseline methods to estimate the reliability degree of ``small" sources accurately in Linked Data, although these methods achieved satisfactory accuracy in certain datasets or applications. 

The experimental results of the average change in the trust value of objects after each iteration are shown in Figure 5.

 \begin{figure}[!htb]
\centering
\includegraphics[width=0.4\textwidth]{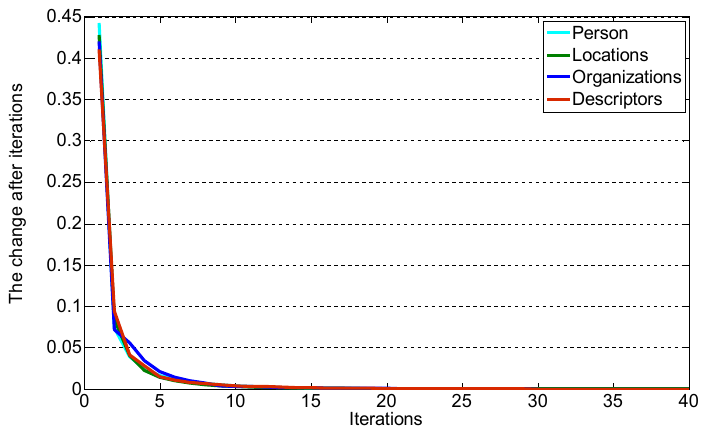}
\caption{Change in the trust values of objects after each iteration}
\end{figure}
It also shows the change decreases rapidly in the first five iterations, and then reaches a stable stage until the convergence criterion is satisfied.
\section{Conclusions}
We have developed an easy-to-use system called TruthDiscover\footnote{Demonstration is available online at http://123.139.159.38:9218/Truth/}, which leverages the topological properties of the Source Belief Graph and the interdependencies between objects to infer the trustworthiness of sources and the trust values of objects. This system is capable of automatically identifying the truth in massive Linked Data with a scale-free property. The experimental results show that TruthDiscover exhibits satisfactory accuracy. The future extension of TruthDiscover should include improving the performance of identifying the truth. This improvement can be achieved by identifying the copying relations of different sources.
\section{ACKNOWLEDGEMENT}
This work is funded by the National Key Research and Development Program of China (Grant No. 2016YFB1000903), the MOE Research Program for Online Education (Grant No. 2016YB166) and the National Science Foundation of China (Grant Nos. 61370019, 61672418, 61532004, 61532015).

%
\bibliographystyle{abbrv}
\bibliography{vldb_sample}  
%
%
%
\balance
\end{document}